\magnification=1200
\baselineskip=13pt
\overfullrule=0pt
\tolerance=100000
\nopagenumbers

\font\tenbifull=cmmib10 \skewchar\tenbifull='177
\font\tenbimed=cmmib7   \skewchar\tenbimed='177
\font\tenbismall=cmmib5  \skewchar\tenbismall='177
\textfont9=\tenbifull
\scriptfont9=\tenbimed
\scriptscriptfont9=\tenbismall

\mathchardef\alpha="710B
\mathchardef\beta="710C
\mathchardef\gamma="710D
\mathchardef\delta="710E
\mathchardef\epsilon="710F
\mathchardef\zeta="7110
\mathchardef\eta="7111
\mathchardef\theta="7112
\mathchardef\iota="7113
\mathchardef\kappa="7114
\mathchardef\lambda="7115
\mathchardef\mu="7116
\mathchardef\nu="7117
\mathchardef\micron="716F
\mathchardef\xi="7118
\mathchardef\pi="7119
\mathchardef\rho="711A
\mathchardef\sigma="711B
\mathchardef\tau="711C
\mathchardef\upsilon="711D
\mathchardef\phi="711E
\mathchardef\chi="711F
\mathchardef\psi="7120
\mathchardef\omega="7121
\mathchardef\varepsilon="7122
\mathchardef\vartheta="7123
\mathchardef\varphi="7124
\mathchardef\varrho="7125
\mathchardef\varsigma="7126
\mathchardef\varpi="7127

 at 8truept

\

{\hfill \hbox{\vbox{\settabs 1\columns
\+ hep-th/9802200\cr
}}}
\centerline{}
\bigskip
\bigskip
\bigskip
\baselineskip=18pt

\centerline{\bf Lightfront Hamiltonian Structures for the Nonlinear Sigma Model}
\vfill
{\baselineskip=11pt
\centerline{Ashok Das}
\medskip
\medskip
\centerline{Department of Physics and Astronomy}
\centerline{University of Rochester}
\centerline{Rochester, NY 14627 -- USA}
}
\vfill

\centerline{\bf {Abstract}}

\medskip
We derive the Dirac brackets for the $O(N)$ nonlinear sigma model in the lightfront description with and without the constraint. We bring out various subtleties that arise including the fact that anti-periodic boundary condition seems to be preferred.
\medskip

\vfill
\eject
\headline={\hfill\folio}
\pageno=1
\bigskip
\leftline{\bf I. Introduction}
\medskip

The lightfront description of field theories [1] has been quite useful in various studies [2-7]. More recently, there has been a serious attempt at obtaining nonperturbative information about a quantum field theory through the lightfront formalism [8]. St
udies of various classical field theories also simplify quite a lot in the lightfront description. For example, the study of sine-Gordon model, its Backlund transformations as well as the soliton solutions are quite straight forward in this formalism [9].
 Similarly, the nonlinear sigma model has a very simple Lax description in the lightfront description and ($O(3)$ sigma model) can be related to the sine-Gordon model quite trivially in this approach [10]. It is surprising, therefore, that the Hamiltonian
 structures for the general $O(N)$ nonlinear sigma model in the lightfront formalism have not been explicitly worked out so far [11-12].

This may have to do with the various subtleties that arise in the lightfront formalism. First, one has to introduce suitable boundary conditions for consistency of the Hamiltonian description [11,13] and more importantly, the inversion of operators, in su
ch a case, becomes quite tedious in general. In this letter, we will derive the Hamiltonian structures for the $O(N)$ nonlinear sigma model in the lightfront description bringing out various subtleties along the way.
\bigskip
\leftline{\bf II. Sigma Model without Constraints}
\medskip

The $O(N)$ nonlinear sigma model is described by the Lagrangian density
$$
{\cal L}={1\over 2}\partial_\mu q^a \partial^\mu q^a\qquad a=1,2,\dots,N\eqno(1)
$$
with the constraint
$$
q^aq^a=1\eqno(2)
$$
Since we are not interested in the dynamics in the transverse directions, we will restrict ourselves to $1+1$ dimensions so that $\mu=0,1$ and we use the metric, $\eta^{\mu\nu}=(1,-1)$. Formulated this way, the sigma model has a manifest $O(N)$ invariance
.

The constraint $(2)$ can be trivially satisfied by introducing the new coordinates
$$
\eqalign{
q^A=&{2\chi^A\over 1+\chi^B\chi^B}={2\chi^A\over 1+\chi^2}\,,\qquad A=1,2,\dots,N-1\cr
q^N=&{1-\chi^2\over 1+\chi^2}
}\eqno(3)
$$
In terms of these $(N-1)$ unconstrained dynamical variables, the Lagrangian density in (1) takes the form
$$
\eqalign{
{\cal L}=&2{\partial_\mu\chi^A \partial^\mu\chi^A\over(1+\chi^2)^2}\cr
=&8{\partial_\eta\chi^A \partial_\xi\chi^A\over(1+\chi^2)^2}
}\eqno(4)
$$
where we have introduced the lightfront coordinates
$$
\eta={t+x\over2}\,,\qquad \xi={t-x\over2}\eqno(5)
$$
It is worth noting here that, in these coordinates, the dynamical equation takes the form
$$
\partial_\eta\left(\partial_\xi\chi^A\over(1+\chi^2)^2\right)+\partial_\xi\left(\partial_\eta\chi^A\over(1+\chi^2)^2\right)+{4\chi^A\partial_\eta\chi^B \partial_\xi\chi^B\over(1+\chi^2)^3}=0\eqno(6)
$$

Treating $\eta$ as the new time coordinate we obtain
$$
\pi^A={\partial{\cal L}\over\partial\partial_\eta\chi^A}=8{(\partial_\xi\chi^A)\over(1+\chi^2)^2}\eqno(7)
$$
This, therefore, defines the primary constraint of this theory to be [14]
$$
\phi^A=\pi^A-8{(\partial_\xi\chi^A)\over(1+\chi^2)^2}\approx 0\eqno(8)
$$
The canonical Hamiltonian density following from (4) vanishes and, therefore, the primary Hamiltonian density takes the form (with the Lagrange multiplier $u^A$)
$$
{\cal H}_p=u^A\phi^A=u^A\left(\pi^A-8{(\partial_\xi\chi^A)\over(1+\chi^2)^2}\right)\eqno(9)
$$
With the canonical Poisson brackets,
$$
\eqalign{
\{\chi^A(\xi),\chi^B(\xi')\}_{\eta=\eta'}=&0=\{\pi^A(\xi),\pi^B(\xi')\}_{\eta=\eta'}\cr
\{\chi^A(\xi),\pi^B(\xi')\}_{\eta=\eta'}=&\delta^{AB}\delta(\xi'-\xi)
}\eqno(10)
$$
it is straightfoward to show that the evolution of the constraint (8) only leads to a relation on the Lagrange multiplier $u^A$, namely
$$
\eqalign{
\partial_\eta\phi^A=&\{\phi^A(\xi),H_p\}\cr
=&-8\partial_\xi\left({u^A\over (1+\chi^2)^2}\right)-8{(\partial_\xi u^A)\over(1+\chi^2)^2}+32{u^B(\partial_\xi\chi^A\chi^B-\chi^A\partial_\xi\chi^B)\over(1+\chi^2)^3}\approx 0
}\eqno(11)
$$
There is no more constraint that is generated. And we note here that although (11) defines a relation for $u^A$, it does not determine it. The Lagrange multiplier can be determined by going to the Hamiltonian equations which identifies
$$
u^A=\partial_\eta\chi^A\eqno(12)
$$
With this, it is easy to show that relation (11) is nothing other than the dynamical equation in (6). This is important to note because, as we have seen, the primary Hamiltonian simply consists of a constraint which can be put to zero at the end of the Di
rac procedure. the dynamical equations, in this case, are contained in the structure of constraints of the theory.

This analysis shows that (8) defines the only constraint of the theory and that (in what follows, we will ignore $\eta=\eta'$ which is understood)
$$
\eqalign{
\{\phi^A(\xi),\phi^B(\xi')\}=&-8\,\delta^{AB}\left\{\partial_\xi{1\over(1+\chi^2(\xi))^2}+
{1\over(1+\chi^2(\xi))^2}\partial_\xi\right\}\delta(\xi-\xi')\cr
&-32{\left(\chi^A(\partial_\xi\chi^B)-\chi^B(\partial_\xi\chi^A)\right)\over
\left(1+\chi^2(\xi)\right)^3}\delta(\xi-\xi')\cr
&=C^{AB}(\xi,\xi')
}\eqno(13)
$$ 
To define the Dirac brackets, we have to invert this matrix and this can be achieved as follows. Let us assume that, acting on a space of functions, $C^{AB}$ gives
$$
\eqalign{
{\overline f}^A(\xi)
=&\int d\xi'C^{AB}(\xi,\xi')f^B(\xi')\cr
=&-{16\over 1+\chi^2(\xi)}\left\{
\partial_\xi\left({f^A(\xi)\over1+\chi^2(\xi)}\right)+
2{\chi^A(\partial_\xi\chi^B)-\chi^B(\partial_\xi\chi^A)\over
1+\chi^2(\xi)}{f^B(\xi)\over1+\chi^2(\xi)}\right\}
}\eqno(14)
$$
Defining
$$
J^{AB}(\xi)=2{\chi^A(\partial_\xi\chi^B)-\chi^B(\partial_\xi\chi^A)\over
1+\chi^2(\xi)}=-J^{BA}(\xi)\eqno(15)
$$
we have
$$
\left(\delta^{AB}\partial_\xi+J^{AB}(\xi)\right){f^B(\xi)\over1+\chi^2(\xi)}=-{1\over 16}(1+\chi^2(\xi)){\overline f}^A(\xi)\eqno(16)
$$
Defining the Greens function for this equation, which has the form,
$$
G^{AB}(\xi,\xi')=\left(P\left(\hbox{e}^{-\int^\xi_{\xi'}J(x)dx}\right)(I\,\epsilon(\xi-\xi')+K)\right)^{AB}\eqno(17)
$$
where $K$ is a matrix independent of the coordinates $\xi$ and $\xi'$. We can write the solution of (16) as
$$
f^A(\xi)=-{1\over 16}\int d\xi'\,(1+\chi^2(\xi))\,G^{AB}(\xi,\xi')\,(1+\chi^2(\xi'))\,{\overline f}^B(\xi')\eqno(18)
$$

The arbitrariness in the Greens function arises primarily because we have not yet specified any boundary condition [11,13] which is so vital to a lightfront description. Conventionally, in dealing with a scalar field theory in the lightfront description, 
one assumes $-L\le\xi\le L$ with $L\to\infty$ at the end of the calculation. Furthermore, one knows that for a simple scalar field theory, both periodic/antiperiodic boundary conditions
$$
\chi^A(L,\eta)=\pm\chi^A(-L,\eta)\eqno(19)
$$
lead to the same result. In this case, however, we will show that the consistency of the procedure would pick out the anti-periodic boundary condition.

The inverse of an operator or the Greens function is defined only on a space of functions which satisfy the same boundary condition as the dynamical variables. Thus, for the anti-periodic boundary condition for the fields, we must have
$$
f^A(L,\eta)=-f^A(-L,\eta)\eqno(20)
$$
which then determines $K$ in (18) to be
$$
K^{AB}=\left(\tanh\left({1\over2}\int_{-L}^{+L}A(x)dx\right)\right)^{AB}
\to\left(\tanh\left({1\over2}\int_{-\infty}^{+\infty}A(x)dx\right)\right)^{AB}\eqno(21)
$$
Such a term, in fact, was already derived [12] in the study of the $O(3)$ sigma model by imposing Jacobi identity (see also [15]). On the other hand, for periodic boundary condition for the fields, we have
$$
f^A(L,\eta)=f^A(-L,\eta)\eqno(22)
$$
which in turn gives
$$
K^{AB}=\left(\coth\left({1\over2}\int_{-L}^{+L}A(x)dx\right)\right)^{AB}
\to\left(\coth\left({1\over2}\int_{-\infty}^{+\infty}A(x)dx\right)\right)^{AB}\eqno(23)
$$
This shows that $K$ and, therefore, the Greens function is uniquely determined depending on the boundary condition used.

Going back to (18), we note from the definition in (14) that we can write
$$
f^A(\xi)=\int d\xi'\,C^{-1}_{AB}(\xi,\xi')\,{\overline f}^B(\xi')\eqno(24)
$$
This determines the inverse matrix from (18) to be
$$
C^{-1}_{AB}(\xi,\xi')=-{1\over 16}(1+\chi^2(\xi))\,G^{AB}(\xi,\xi')\,(1+\chi^2(\xi'))\eqno(25)
$$
This is unique once we impose the boundary conditions and is antisymmetric as it should be. The Dirac brackets can now be defined. The only fundamental Dirac bracket is
$$
\eqalign{
\{\chi^A(\xi),\chi^B(\xi')\}_D=&\{\chi^A(\xi),\chi^B(\xi')\}\cr
&-\int d\xi''\xi'''
\{\chi^A(\xi),\phi^Q(\xi'')\}\,C^{-1}_{PQ}(\xi'',\xi''')\,
\{\phi^Q(\xi'''),\chi^B(\xi')\}\cr
=&-{1\over 16}(1+\chi^2(\xi))\,G^{AB}(\xi,\xi')\,(1+\chi^2(\xi'))
}\eqno(26)
$$

We note at this point that if we restrict ourselves to $O(2)$ for which $A,B=1$, $J^{AB}=0$ and consequently, for the antiperiodic boundary condition, $K=0$. The Hamiltonian structure has the form
$$
{\cal D}=-{1\over 16}(1+\chi^2)\,\partial^{-1}\,(1+\chi^2)\eqno(27)
$$
Using the method of prolongation [16-17], one can readily check that this structure satisfies Jacobi identity. For periodic boundary condition, on the other hand, with $J^{AB}=0$, $K\to\infty$ and, consequently, the Hamiltonian structure is ill defined. T
his, therefore, seems to pick out the anti-periodic boundary condition for consistency and that is what  we will use in the rest of the paper. We emphasize that for the anti-periodic boundary condition, $K$ is given in (21).
\bigskip
\leftline{\bf III. $O(N)$ Invariant Description of the Sigma Model}
\medskip

For an $O(N)$ invariant description, we do not solve the constraint (2) explicitly. Rather, we incorporate it into the Lagrangian density through a Lagrange multiplier as
$$
\eqalign{
{\cal L}=&{1\over 2}\partial_\mu q^a \partial^\mu q^a+{\lambda\over2}(q^aq^a-1)\cr
=&\partial_\eta q^a \partial_\xi q^a+{\lambda\over2}(q^aq^a-1)\qquad a=1,2,\dots,N}\eqno(28)
$$
The primary constraints, in this case, turn out to be
$$
\eqalign{
\pi^a=&{\partial{\cal L}\over\partial\partial_\eta q^a}=\partial_\xi q^a\cr
\pi=&{\partial{\cal L}\over\partial\partial_\eta \lambda}=0
}\eqno(29)
$$
so that we can write
$$
\eqalign{
\phi^a_1=&\pi^a-\partial_\xi q^a\approx 0\cr
\phi_2=&\pi\approx 0
}\eqno(30)
$$
The canonical Hamiltonian density is easily obtained to be
$$
{\cal H}_c=-{\lambda\over2}(q^aq^a-1)
$$
and consequently, the primary Hamiltonian density takes the form
$$
{\cal H}_p=-{\lambda\over2}(q^aq^a-1)+u_1^a\phi_1^a+u_2\phi\eqno(31)
$$

With the canonical Poisson brackets
$$
\eqalign{
\{q^a(\xi),q^b(\xi')\}=&\{\pi^a(\xi),\pi^b(\xi')\}=\{\lambda(\xi),\lambda(\xi')\}=\{\pi(\xi),\pi(\xi')\}=0\cr
\{q^a(\xi),\pi^b(\xi')\}=&\delta^{ab}\delta(\xi-\xi')\cr
\{\lambda(\xi),\pi(\xi')\}=&\delta(\xi-\xi')
}\eqno(32)
$$
as well as the $\lambda$ degrees of freedom in involution with the $q^a$ degrees of freedom, we can examine the evolution of the constraints
$$
\eqalign{
\partial_\eta\phi_1^a=&\{\phi_1^a(\xi),H_p\}=-2\left(\partial_\xi u_1^a(\xi)\right)+\lambda q^a(\xi)\approx 0\cr
\partial_\eta\phi=&\{\phi(\xi),H_p\}={1\over2}(q^aq^a-1)\approx0\cr
}\eqno(33)
$$
While the first equation in (33) gives a relation for the Lagrange multiplier $u_1^a$, the second generates a true secondary constraint
$$
\phi_3=q^aq^a-1\approx 0\eqno(34)
$$
Furthermore,
$$
\partial_\eta\phi_3=\{\phi_3(\xi),H_p\}=2q^a(\xi)\,u_1^a(\xi)\approx 0\eqno(35)
$$
Consequently, no more constraints are generated. In passing we note that one can determine the Lagrange multipliers $u_1^a$ and $u_2$ from the Hamiltonian equations and they show that the dynamical equations are contained in the constraint relations in (3
3)-(35). This is like the previous section because upon putting the constraints strongly to zero at the end, the primary Hamiltonian also vanishes.

There are three constraints in the theory, namely (30) and (34). Of these, $\phi_2$ is trivially seen to be first class. Consequently, we can choose a gauge
$$
\phi_4=\lambda-c\approx 0\eqno(36)
$$
where $c$ is a constant. Without going into detail, we note here that different
values of $c$ correspond to choosing different normalizations for the derivatives of the coordinates $q^a$. The set of constraints $(\phi_2,\phi_4)$ now becomes second class and modify the $\lambda$ brackets such that
$$
\{\lambda(\xi),\lambda(\xi')\}_D=\{\pi(\xi),\pi(\xi')\}_D=
\{\lambda(\xi),\pi(\xi')\}_D=0
\eqno(37)
$$ 
The $\lambda$-degrees of freedom can, therefore, be ignored.

The remaining two constraints have the following nontrivial Poisson Bracket structure
$$
\eqalign{
\{\phi_1^a(\xi),\phi_1^b(\xi')\}=&-2\delta^{ab}\partial_\xi\delta(\xi-\xi')\cr
\{\phi_1^a(\xi),\phi_3(\xi')\}=&-2q^a(\xi)\,\delta(\xi-\xi')=-\{\phi_3(\xi),\phi_1^a(\xi')\}
}\eqno(38)
$$
Therefore, defining the matrix of the Poisson brackets as
$$
C^{\alpha\beta}(\xi,\xi')=\pmatrix{-2\delta^{ab}\partial_\xi &-2q^a\cr
\noalign{\vskip .3truecm}%
2q^b &0}\delta(\xi-\xi')\qquad\alpha,\beta=a,3\eqno(39)
$$
we note that the inverse of this matrix will define the Dirac brackets. To obtain the inverse, we again examine the action of $C^{\alpha\beta}$ on a space of matrix functions such that
$$
\pmatrix{{\overline f}^a(\xi)\cr
\noalign{\vskip .3truecm}%
{\overline F}(\xi)}=\int d\xi'\,
\pmatrix{C^{ab}(\xi,\xi') &C^{a3}(\xi,\xi')\cr
\noalign{\vskip .3truecm}%
C^{3b}(\xi,\xi')&C^{33}(\xi,\xi')}
\pmatrix{{f}^b(\xi')\cr
\noalign{\vskip .3truecm}%
{F}(\xi')}\eqno(40)
$$
Explicitly, this gives
$$
\eqalign{
{\overline F}(\xi)=&2q^af^a(\xi)\cr
{\overline f}^a(\xi)=&-2(\partial_\xi f^a)-2q^aF(\xi)
}\eqno(41)
$$
The first of this simply expresses the longitudinal component of $f^a(\xi)$ as (longitudinal with respect to $q^a$)
$$
q^af^a(\xi)={1\over2}{\overline F}(\xi)\eqno(42)
$$
The second equation in (41) can be decomposed into longitudinal and transverse components with respect to $q^a$ and gives respectively
$$
\eqalignno{
F(\xi)=&-{q^a\over q^2}\left((\partial_\xi f^a)+{1\over2}{\overline f}^a(\xi)\right)&(43)\cr
\partial_\xi f^a_T+J^{ab}f^b_T=&-{1\over2}\left(\delta^{ab}-{q^aq^b\over q^2}\right)\left({\overline f}^b+{(\partial_\xi q^b)\over q^2}{\overline F}\right)
&(44)}
$$
where we have defined
$$
\eqalign{
f^a_T(\xi)=&f^a-{q^a\over q^2}(q\cdot f)=f^a-{1\over2}{q^a\over q^2}{\overline F}\cr
J^{ab}(\xi)=&{q^a(\partial_\xi q^b)-q^b(\partial_\xi q^a)\over q^2}=-J^{ba}(\xi)
}\eqno(45)
$$

Once again, we can solve Eq. (44) by the method of Greens functions (restricting to space of transverse functions with anti-periodic boundary conditions) to give
$$
f^a_T(\xi)=-{1\over2}\int d\xi'\,\left(\delta^{ap}-{q^aq^p\over q^2}(\xi)\right)\,D^{pr}(\xi,\xi')\,\left(\delta^{rb}-{q^rq^b\over q^2}(\xi')\right)\,{\overline f}^b(\xi')\eqno(46)
$$
where, as before,
$$
D^{pr}(\xi,\xi')=\left(P\hbox{e}^{-\int^\xi_{\xi'}A(x)dx}(I\,\epsilon(\xi-\xi')+K)\right)^{pr}\eqno(47)
$$
with $K$ given in (21). Using (43)-(46), we can write
$$
\eqalign{
f^a(\xi)=&-{1\over2}\int d\xi'\,\left[G^{ab}(\xi,\xi'){\overline f}^b(\xi')
+\left\{-{q^a\over q^2}\delta(\xi-\xi')+G^{ab}(\xi,\xi'){(\partial_{\xi'} q^b)\over q^2(\xi')}\right\}{\overline F}(\xi')\right]\cr
F(\xi)=&-{1\over2}\int d\xi'\,\left[\left\{{q^a\over q^2}\delta(\xi-\xi')+{(\partial_{\xi} q^a)\over q^2(\xi)}G^{ab}(\xi,\xi')\right\}{\overline f}^b(\xi')\right.\cr
&+\left.\left\{{1\over 2}\left(\partial_\xi{1\over q^2(\xi)}+{1\over q^2(\xi)}\partial_\xi\right)\delta(\xi-\xi')+{(\partial_{\xi} q^a)\over q^2(\xi)}G^{ab}(\xi,\xi'){(\partial_{\xi'} q^b)\over q^2(\xi')}\right\}{\overline F}(\xi')\right]
}\eqno(48)
$$
Here we have defined, for simplicity
$$
G^{ab}(\xi,\xi')=\left(\delta^{ap}-{q^aq^p\over q^2}(\xi)\right)\,D^{pr}(\xi,\xi')\,\left(\delta^{rb}-{q^rq^b\over q^2}(\xi')\right)\eqno(49)
$$

Thus, we see from (48) that we can invert the relation in (40) and write
$$
\pmatrix{{f}^a(\xi)\cr
\noalign{\vskip .3truecm}%
{F}(\xi)}=\int d\xi'\,
\pmatrix{C^{-1}_{ab}(\xi,\xi') &C^{-1}_{a3}(\xi,\xi')\cr
\noalign{\vskip .3truecm}%
C^{-1}_{3b}(\xi,\xi')&C^{-1}_{33}(\xi,\xi')}
\pmatrix{{\overline f}^b(\xi')\cr
\noalign{\vskip .3truecm}%
{\overline F}(\xi')}\eqno(50)
$$
with the matrix elements for the inverse given by
$$
\eqalign{
C^{-1}_{ab}(\xi,\xi')=&-{1\over2}G^{ab}(\xi,\xi')\cr
C^{-1}_{a3}(\xi,\xi')=&-{1\over2}\left[-{q^a\over q^2}\delta(\xi-\xi')+
G^{ab}(\xi,\xi'){(\partial_{\xi'}q^b)\over q^2(\xi')}\right]\cr
C^{-1}_{3a}(\xi,\xi')=&-{1\over2}\left[{q^a\over q^2}\delta(\xi-\xi')+
{(\partial_{\xi}q^b)\over q^2(\xi)}G^{ba}(\xi,\xi')\right]\cr
C^{-1}_{33}(\xi,\xi')=&-{1\over2}\left[{1\over2}\left(\partial_\xi{1\over q^2(\xi)}+{1\over q^2(\xi)}\partial_\xi\right)\delta(\xi-\xi')+
{(\partial_{\xi}q^a)\over q^2(\xi)}G^{ab}(\xi,\xi'){(\partial_{\xi'}q^b)\over q^2(\xi')}\right]\cr
}\eqno(51)
$$
The inverse matrix has the necessary antisymmetry properties. Furthermore, it an be used to define the Dirac brackets of the theory. The only fundamental bracket has the form
$$
\eqalign{
\{q^a(\xi),q^b(\xi')\}_D=&\{q^a(\xi),q^b(\xi')\}\cr
&-\int d\xi''\xi'''
\{q^a(\xi),\phi_1^p(\xi'')\}\,C^{-1}_{pr}(\xi'',\xi''')\,
\{\phi_1^r(\xi'''),q^b(\xi')\}\cr
=&-{1\over 2}G^{ab}(\xi,\xi')
}\eqno(52)
$$
All other brackets can be derived from these.
\bigskip
\leftline{\bf IV. Conclusions}
\medskip
 
We have derived the consistent Dirac brackets for the $O(N)$ nonlinear sigma model in the lightfront description. Surprisingly, it seems to prefer anti-periodic boundary conditions for the scalar fields as opposed to the periodic boundary conditions. Vari
ous subtleties that arise have also been explained in detail.

\bigskip
\leftline{\bf Acknowledgments}
\medskip

I would like to thank Prof. J. C. Brunelli for many interesting discussions and the members of Departamento de F\'\i sica da Universidade Federal de Santa Catarina, Brazil,  for their hospitality during the period when this work was done. It is a pleasure
 to thank Sue Brightman, Judy Mack and Dr. Wolfgang Scherer for some essential help towards the completion of this project.  This work  is supported in part by the U.S. Department of Energy Grant No. DE-FG-02-91ER40685, NSF-INT-9602559 as well as a Fulbri
ght grant.  

Note added: Since writing this paper, I have become aware of the fact that the question of the light front Dirac quantization for the scalar field theory, with a careful analysis of the boundary conditions, was first carried out in [18] (see also [19]).

\medskip
\vfill\eject
\leftline{\bf References}
\bigskip

\item{1.}{P. A. M. Dirac, Rev. Mod. Phys. {\bf 21} (1949) 392.}
\item{2.}{S. Fubini and G. Furlan, Physics {\bf 1} (1965) 229; S. Weinberg, Phys. Rev. {\bf 150} (1966) 1313; J. Jersak and J. Stern, Nucl. Phys. {\bf B7} (1968) 413; H. Leutwyler in Springer Tracks in Modern Physics, vol. 50, ed. G. H\"ohler, Springer, B
erlin 1969.}
\item{3.}{J. D. Bjorken, Phys. Rev. {\bf 179} (1969) 1547; S. D. Drell, D. Levy and T. M. Yan, Phys. Rev {\bf 187} (1969) 2159, {\bf D1} (1970) 1035.}
\item{4.}{R. Jackiw in Springer Tracks in Modern Physics, ed G. H\"ohler, Springer, Berlin 1972.}
\item{5.}{S. J. Chang and S. K. Ma, Phys Rev. {\bf 180} (1969) 1506; J. B. Kogut and D. E. Soper, Phys. Rev {\bf D1} (1970) 2901; J. D. Bjorken, J. B. Kogut and D. E. Soper, Phys. Rev. {\bf D3} (1971) 1382; R. A. Neville and F. Rohrlich, Phys. Rev. {\bf D
3} (1971) 1692.}
\item{6.}{S. J. Chang, R. G. Root and T. M. Yan, Phys. Rev. {\bf D7} (1973) 1133; S. J. Chang and T. M. Yan, Phys. Rev. {\bf D7} (1973) 1147; T. M. Yan, Phys. Rev. {\bf D7} (1973) 1760, 1780.}
\item{7.}{H. C. Pauli and S. J. Brodsky, Phys. Rev. {\bf D32} (1985) 1993, 2001; T. Eller, H. C. Pauli and S. J. Brodsky, Phys. Rev. {\bf D35} (1987) 1439; 
H. C. Pauli and S. J. Brodsky, hep-ph/9705477.}
\item{8.}{R. J. Perry, A. Harindranath and K. G. Wilson, Phys. Rev. Lett. {\bf 65} (1990) 2959; R. J. Perry and A. Harindranath, Phys. Rev. {\bf D43} (1991) 492, 4051; D. Mustaki, S. Pinsky, J. Shigemitsu and K. G. Wilson, Phys. Rev. {\bf D43} (1991) 3411
.}
\item{9.}{See, for example, A. Das ``Integrable Models'', World Scientific, Singapore 1989.}
\item{10.}{K. Pohlmeyer, Comm. Math. Phys. {\bf 46} (1976) 207; V. E. Zakharov and A. V. Mikhailov, Sov. Phys. JETP {\bf 47} (1978) 1017.}
\item{11.}{P. J. Steinhard, Ann. Phys. {\bf 128} (1980) 425. Here the discussion is for the $O(2)$ nonlinear sigma model.}
\item{12.}{S. B. Pereslegin and V. A. Franke, Theo. Math. Phys. {\bf 59} (1984) 365. These authors discuss the Hamiltonian structure for the $O(3)$ model. While the first part of the paper gives essentially the correct Hamiltonian structure, the second pa
rt contains the incorrect matrix inverse leading to Dirac brackets that do not satisfy the constraints.}
\item{13.}{T. Heinzl, S. Krusche and E. Werner, Phys. Lett. {\bf B256} (1991) 55; T. Heinzl and E. Werner, Phys. Lett. {\bf B275} (1992) 410; T. Heinzl and E. Werner, Z. Phys. {\bf C62} (1994) 521}
\item{14.}{P. A. M. Dirac, Can. J. Math. {\bf 1} (1950) 1; P. A. Dirac, ``Lectures in Quantum Mechanics'', Yeshiva University, New York 1964; A. Hanson, T. Regge and C. Teitelboim, ``Constrained Hamiltonian Systems'', Acad. Naz. dei  Lincei, Rome 1976; K
. Sundermeyer, ``Constrained Dynamics'', Lecture Notes in Physcis, {\bf 169}, Springer, Berlin 1982.}
\item{15.}{In ref. 12, there is an arbitrary parameter $\kappa$ which is not allowed by the boundary conditions.}
\item{16.}{P. J. Olver, ``Applications of Lie Groups to Differential Equations'', Springer, Berlin 1986.}
\item{17.}{J. C. Brunelli and A. Das, Int. J. Mod. Phys. {\bf A10} (1995) 4563.}
\item{18.}{T. Maskawa and K. Yamawaki, Prog. Theo. Phys. {\bf 56} (1976) 270.}
\item{19.}{N. Nakanishi and K. Yamawaki, Nuc. Phys. {\bf B122} (1977) 15; Y. Kim, S. Tsujimaru and K. Yamawaki, Phys. Rev. Lett. {\bf 74} (1995) 4771; K. Yamawaki, hep-th/9802037.}
\end